\documentclass[aps, prl, reprint, twocolumn, superscriptaddress]{revtex4-1}
\usepackage{lipsum}
\usepackage{multirow}
\usepackage{graphicx}
\usepackage{longtable}
\usepackage[utf8]{inputenc}
\usepackage[T1]{fontenc}
\usepackage{epstopdf}
\usepackage{textcomp}
\usepackage[cmyk,dvipsnames]{xcolor} % For obvious reasons
\usepackage{amsbsy}
\usepackage{amsmath}
\usepackage{amssymb}
\usepackage{amsfonts}
\usepackage{dsfont}
\usepackage{mathtools}
\usepackage{braket}
\usepackage{array}
\usepackage{gensymb}
\usepackage{placeins}
\usepackage{dashrule}
\usepackage{xcolor}
\usepackage{booktabs}
\usepackage{hyperref}
\usepackage{float}
\usepackage{braket}
\usepackage{ulem}

\definecolor{ao(english)}{rgb}{0.0, 1.0, 0.0}

\definecolor{pacificb}{HTML}{1CA9C9}

\makeatletter
  \def\my@tag@font{\normalsize}
  \def\maketag@@@#1{\hbox{\m@th\normalfont\my@tag@font#1}}
  \let\amsmath@eqref\eqref
  \renewcommand\eqref[1]{{\let\my@tag@font\relax\amsmath@eqref{#1}}}
\makeatother

\allowdisplaybreaks

\begin{document}
% Thermal generation of chiral droplets in chiral magnets
\title{
%\NSK{Thermally assisted chiral droplet nucleation}
%Thermally induced nucleation of chiral droplet\\
Thermal generation of droplet soliton in chiral magnet
}

\author{Vladyslav~M.~Kuchkin}
 \email{v.kuchkin@fz-juelich.de}
 \affiliation{Peter Gr\"unberg Institute and Institute for Advanced Simulation, Forschungszentrum J\"ulich and JARA, 52425 J\"ulich, Germany}
 \affiliation{Department of Physics, RWTH Aachen University, 52056 Aachen, Germany}

\author{Pavel~F.~Bessarab}
%\affiliation{Peter Gr\"unberg Institute and Institute for Advanced Simulation, Forschungszentrum J\"ulich and JARA, 52425 J\"ulich, Germany}
\affiliation{Science Institute of the University of Iceland, 107 Reykjav\'ik, Iceland}
\affiliation{ITMO University, 197101 St. Petersburg, Russia}

\author{Nikolai~S.~Kiselev}
 %\email{n.kiselev@fz-juelich.de}
 \affiliation{Peter Gr\"unberg Institute and Institute for Advanced Simulation, Forschungszentrum J\"ulich and JARA, 52425 J\"ulich, Germany}

\date{\today}

\begin{abstract}%152
Controlled creation of localized magnetic textures beyond conventional $\pi$-skyrmions is an important problem in the field of magnetism. Here by means of spin dynamics simulations, Monte Carlo simulations and harmonic transition state theory we demonstrate that an elementary chiral magnetic soliton with zero topological charge -- the chiral droplet -- can be reliably created  by thermal fluctuations in the presence of the tilted magnetic field. The proposed protocol relies on an unusual kinetics combining the effects of the entropic stabilization and low energy barrier for the nucleation of a topologically-trivial state.
Following this protocol by varying temperature and the tilt of the external magnetic field one can selectively generate chiral droplets or $\pi$-skyrmions in a single system.
The coexistence of two distinct
magnetic solitons establishes a basis for a rich magnetization dynamics and opens up the possibility for the construction of more complex magnetic textures such as skyrmion bags and skyrmions with chiral kinks.
\end{abstract}

\maketitle 
% \section{Introduction}

%128
The model of chiral magnets 
allows surprisingly many spatially localized statically stable solutions. 
Together with originally 
reported solutions, also known as $k\pi$-skyrmions (Sks)~\cite{Bogdanov_99}
a large diversity of non-axially symmetric solitons with an arbitrary topological index 
has recently been discovered in the two-dimensional (2D) model of chiral magnet. 
The latter include skyrmion bags~\cite{Rybakov_19,Foster_19} and Sks with chiral kinks (CKs)~\cite{Cheng_19,Kuchkin_20i,Kuchkin_20ii,Kuchkin_21}. 
Recently, the direct observation of skyrmion bags by means of Lorentz transmission electron microscopy and their current-induced motion have been reported in Ref.~\cite{Tang_21}.
The experimental evidence for  CKs has been provided in Ref.~\cite{Li_20}.
Co-existence of various types of solitons in a single system is fundamentally interesting and technologically appealing. 
However, since the localized states beyond conventional Sks are typically metastable states, their controllable nucleation is challenging.

%136
Here we suggest a reliable protocol for generating an elementary magnetic soliton containing a single CK -- the chiral  droplet (CD), also referred to as a chimera skyrmion~\cite{rozsa_2017} -- by means of thermal fluctuations and oblique magnetic field. 
We refer to CD as an elementary chiral soliton because it is the most compact non-axially symmetric soliton containing only one CK. 
The CD texture has previously been reported as a statically stable solution~\cite{Kuchkin_20ii,rozsa_2017} and a transient state during the asymmetric Sk collapse~\cite{muckel_21,Meyer_19}.
In contrast to $k\pi$-skyrmions, the interparticle interaction potentials for CD with other solitons are a strongly asymmetric due to the presence of the CK~\cite{Kuchkin_20i}. 
As a consequence, CDs may attract or repel other solitons depending on their mutual orientation. 
This provides a basis for the skyrmion fusion, and, thereby, creation of more complex magnetic textures.

%140
Although CDs represent excitations in the ferromagnetic (FM) background, their large entropy enables entropic stabilization, similar to what was reported for conventional $\pi$-skyrmions~\cite{desplat_2018,malottki_2019,varentcova_2020}. 
On the other hand, CDs belong to a class of topologically trivial solitons~\cite{Kuchkin_20ii}.
Because of that one may expect lower energy barriers for their nucleation compared to that for topologically nontrivial textures. 
As a result, there are prerequisites for an effective thermal generation of CDs. 
Indeed, we found that under tilted magnetic field and moderate thermal fluctuations, the spontaneous nucleation of CDs dominates the $\pi$-Sk nucleation by several orders of magnitude. 
Noteworthy, by varying the temperature and the tilt angle of the external field one can selectively nucleate either $\pi$-Sks or CDs. 
These findings are supported by the consistency of stochastic Landau-Lifshitz-Gilbert (LLG) simulations, Monte Carlo simulations, and analysis based on the transition state theory.

%10
We consider a classical spin Hamiltonian on a square lattice:
\begin{align}
E =\!-J\!\sum_{\left\langle i,j\right\rangle}\!\mathbf{n}_{i}\cdot\mathbf{n}_{j} - \sum_{\left\langle i,j\right\rangle}\!\mathbf{D}_{ij}\cdot[\mathbf{n}_{i}\!\times\!\mathbf{n}_{j}] - \mu_\mathrm{s}\mathbf{B}\sum_{i}\mathbf{n}_{i},
\label{Hamiltonian}
\end{align}
%94
were $\mathbf{n}_i$ is the normalized magnetization vector at lattice site $i$, $J$ and $\mathbf{D}=D\hat{\mathbf{r}}_{ij}$ are the Heisenberg exchange constant and Dzyaloshinskii-Moriya (DM) vector, respectively, $\hat{\mathbf{r}}_{ij}$ is the unit vector between sites $i$ and $j$, $\mu_\mathrm{s}$ is the magnitude of the magnetic moment at each site, and 
$\mathbf{B}$ is the external magnetic field.
The symbol $\left\langle i,j\right\rangle$ denotes summation over unique nearest neighbor pairs.
The ratio between $J$ and $D$ defines the equilibrium period of helical spin spirals, $L_\mathrm{D}=2\pi Ja/D$, with $a$ being the lattice constant, and characteristic magnetic field, $B_\mathrm{D}=D^{2}/(J \mu_\mathrm{s})$.

% caption 84
\begin{figure*}[ht]
\centering
\includegraphics[width=17cm]{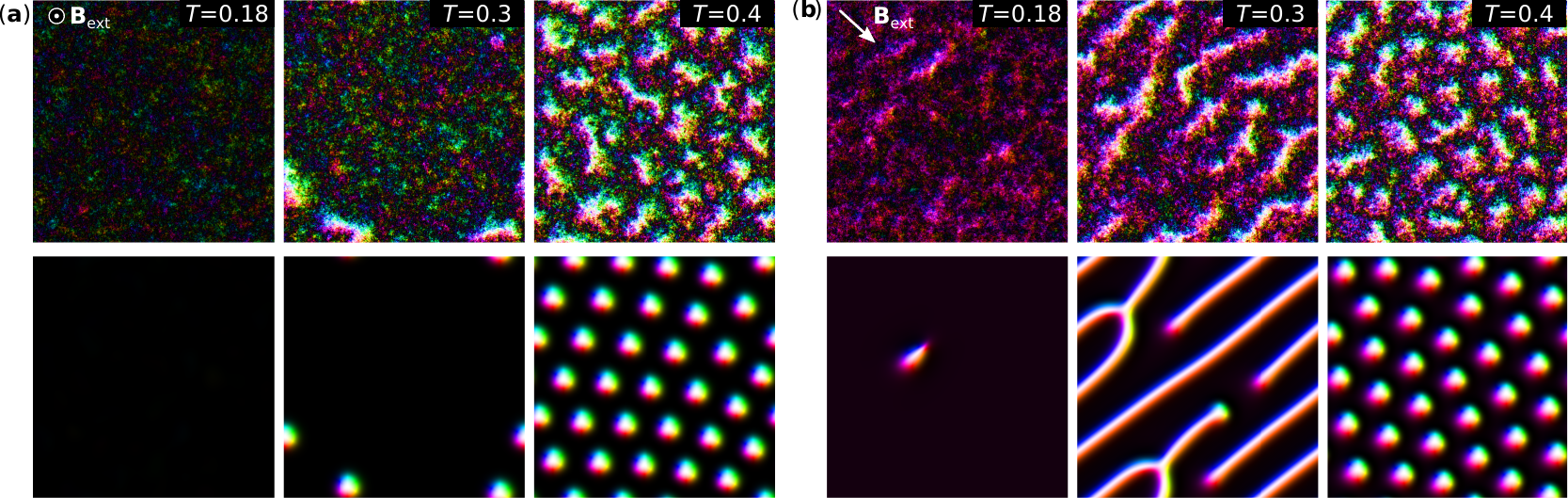}
\caption{\small 
\textbf{(a)} Illustrates the case of perpendicular magnetic field, $h=0.643$, $\vartheta=0$ and 
\textbf{(b)} corresponds to the tilted magnetic field, $h=0.645$, $\vartheta=0.4$. 
The simulations were performed on a square domain, $L_\mathrm{x}\!=\!L_\mathrm{y}\!=\!8L_\mathrm{D}$, with periodic boundary conditions in the $xy$-plane, and $\mathbf{n}(\mathbf{r})||\mathbf{B}_\mathrm{ext}$ in the initial state.
The top row of images represents the snapshots of the system at different temperatures taken at thermal equilibrium after $\sim 10^6$ LLG iterations. Each image in the bottom raw corresponds to the top image after setting $T=0$ and energy relaxation.}
\label{Fig1}
\end{figure*}

We consider the case when magnetic filed is tilted with respect to the plane normal,  $\mathbf{h}=\mathbf{B}/B_\mathrm{D}=h(\sin\vartheta\cos\varphi, \sin\vartheta\sin\varphi, \cos\vartheta)$ and parametrized by the polar angle $\vartheta$  and azimuthal angle $\varphi$.
For the parameters of $J$ and $D$ used in our simulations and providing a relatively large $L_\mathrm{D}=64a$, the Hamiltonian  \eqref{Hamiltonian} becomes nearly isotropic in the $xy$-plane. 
In this case, the choice of angle $\varphi$  does not affect the results, but for definiteness we fix $\varphi=-\pi/4$.

We simulate spin dynamics at finite temperature using the stochastic LLG equation:
\begin{align}
 & \dfrac{\partial\mathbf{n}_{i}}{\partial t}=-\mathbf{n}_i\times\left(\mathbf{B}_\mathrm{eff}^{i}+\mathbf{B}_\mathrm{fluc}^{i}\right)+\alpha\mathbf{n}_{i}\times\frac{\partial\mathbf{n}_{i}}{\partial t},\label{LLG}
\end{align}
%103
where $t$ is a dimensionless time scaled
by $J\gamma \mu_\mathrm{s}^{-1}$, with $\gamma$ being the gyromagnetic ratio,
$\alpha$ is the Gilbert damping parameter, $\mathbf{B}_\mathrm{eff}^{i}=-\dfrac1J\dfrac{\partial E}{\partial \mathbf{n}_{i}}$ is a dimensionless effective field and  $\mathbf{B}_\mathrm{fluc}^{i}$ is the the fluctuating field representing uncorrelated Gaussian white noise with correlation coefficient proportional to temperature, $T$.
For the numerical integration of Eq.~\eqref{LLG}, we use the semi-implicit method provided in Ref.~\cite{Mentink_10} assuming $\alpha=0.3$ and time step $\Delta t=0.01$.
For the chosen coupling parameters we estimate the critical temperature, $T_\mathrm{c}\simeq 0.7 J/{k_\mathrm{B}}$ (see Ref.~\cite{suppl}).
For the results presented below, the temperature is always $T<T_\mathrm{c}$ and given in units of $J/{k_\mathrm{B}}$.

%183
The top row of images in Fig.~\ref{Fig1} provides representative snapshots of the LLG simulations showing the system at various temperatures and different tilt angles of the external magnetic field, $\vartheta=0$ in \textbf{a} and $\vartheta=0.4$ in \textbf{b}. 
To make the presence of the localized magnetic textures in the system more evident, in the bottom row of images we provide corresponding snapshots of the system after cooling by setting $T=0$ in \eqref{LLG}.
In the case of perpendicular magnetic field, $\vartheta=0$ we observe spontaneous nucleation of $\pi$-Sks only, while at tilted magnetic field, $\vartheta=0.4$, we observe the nucleation of CDs. 
Noticeably, the temperature required for the nucleation of CD is significantly lower than that for the $\pi$-Sk nucleation, compare the snapshots in \textbf{a} and \textbf{b} at $T\!=\!0.18 \,J/k_\mathrm{B}$.  
Thus, applying the tilted field and varying the temperature one can selectively nucleate either CDs or $\pi$-Sks. 
At high temperature, however, in both cases, we observe emergence of $\pi$-Sks which in appropriate range of fields tend to form a regular lattice.
There is a critical tilt angle, $\vartheta_\mathrm{c}\approx 50^\circ$, above which the $\pi$-Sk lattice becomes unstable~\cite{Leonov_17}.

%59 
The CD orientation in 2D space can be defined by the in-plane component of the net magnetization of the CD, $\mathbf{m}=\sum_{i} \mathbf{n}_{i}-\mathbf{n}_0$, where $\mathbf{n}_0$ is the magnetization far from the soliton $\mathbf{n}_0=\mathbf{n}(\mathbf{r})$, for $r\rightarrow \infty$.
In the presence of a tilted magnetic field, vector $\mathbf{m}$ of CD is always parallel to the in-plane projection of the external field, $\mathbf{h}$.

%caption 102
\begin{figure*}%[h!]
\centering
\includegraphics[width=17cm]{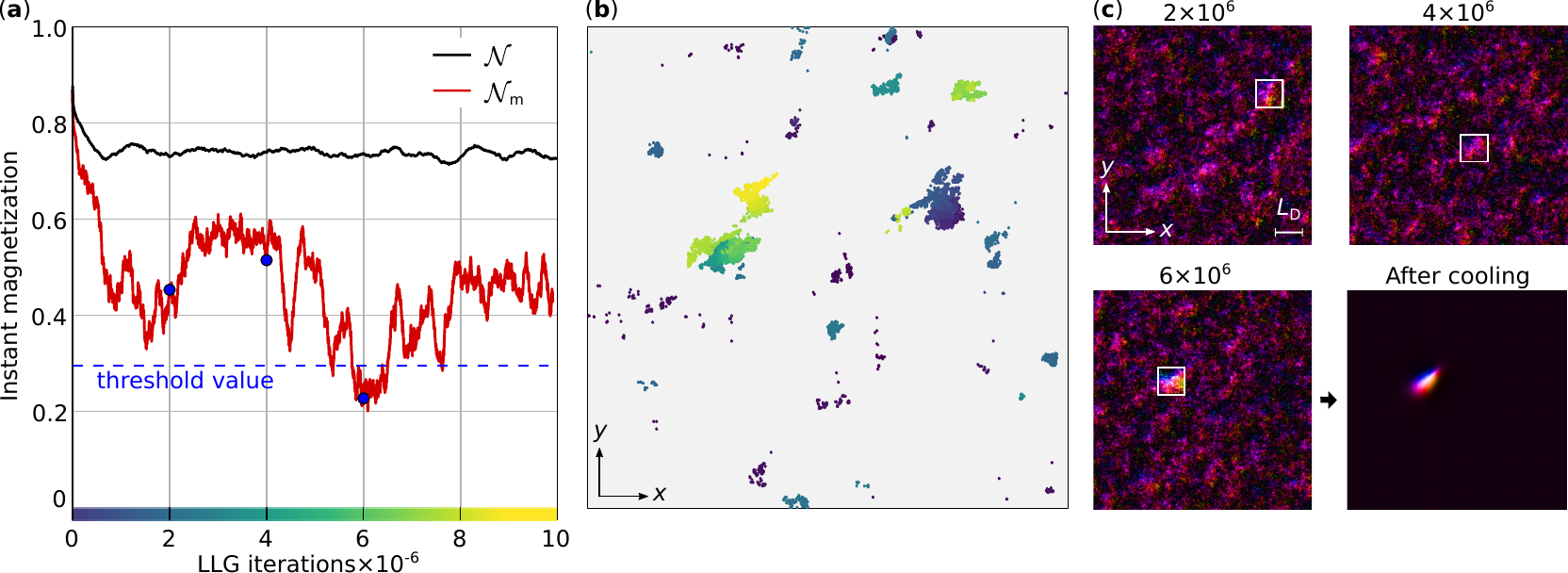}
\caption{\small
\textbf{(a)} shows the evolution of the average magnetization, $\mathcal{N}$, (black) and its maximal value, $\mathcal{N}_\mathrm{m}$, (red) at temperature $T=0.18$, as obtained in the LLG simulations. The dashed blue line corresponds to the threshold value of $0.3$ for $\mathcal{N}_\mathrm{m}$. The position of sub-domain $\Omega_\mathrm{m}$ corresponding to $\mathcal{N}_\mathrm{m}$ is given in \textbf{(b)}, the colors encode the time [see the color bar in \textbf{(a)}]. Blue dots in \textbf{(a)} mark instants of time for which magnetic textures are shown in \textbf{(c)}. The white squares of size $L_\mathrm{D}\!\times \!L_\mathrm{D}$ correspond to the sub-domains $\Omega_\mathrm{m}$. The right bottom image in \textbf{(c)} is obtained after relaxation at $T=0$.}
\label{Fig2}
\end{figure*}

%76
Figure~\ref{Fig2} illustrates the details of the CD nucleation process, as obtained from the LLG simulations. 
The black curve in Fig.\ref{Fig2}~\textbf{a} shows an averaged out-of-plane component of magnetization, $\mathcal{N}=\sum_{i} n_{i,\mathrm{z}}/N$ where $i$ runs over all $N$ spins.
With time, $\mathcal{N}$ converges to its equilibrium value, which for chosen parameters equals $0.74$. 
The Monte Carlo (MC) simulations~\cite{suppl} are fully consistent with the results of LLG simulations and show similar behavior of $\mathcal{N}$ in reaching thermal equilibrium.

%201
$\pi$-Sks are characterized by topological charge $Q=-1$, hence the event of their nucleation can be identified by calculation of $Q$. 
In contrast, the CDs have zero topological charge, which makes this approach not applicable. 
To identify the presence of CDs in the system we employ an alternative method based on time tracing of the net magnetization, as described in the following. 
We split the whole simulated domain into overlapping sub-domains $\Omega_j$ of fixed $L_\mathrm{D}\times L_\mathrm{D}$-size containing $N^\prime=L_\mathrm{D}^2/a^2$ spins.
Taking into account the periodic boundary conditions, the total number of such sub-domains equals the number of spins in the system, $N$.
At each time step, we calculate the averaged out-of-plane magnetization for each sub-domain,
$\mathcal{N}_j=\sum_{i}n_{i,\mathrm{z}}/N^\prime$, $i\in \Omega_j$.
Then, we identify the sub-domain $\Omega_\mathrm{m}$ with minimal $\mathcal{N}_j$ denoted $\mathcal{N}_\mathrm{m}$. 
The red curve in Fig.\ref{Fig2}~\textbf{a} shows representative dependency of 
$\mathcal{N}_\mathrm{m}$ on the LLG simulation step. 
The event of soliton nucleation is signaled by a drop of  $\mathcal{N}_\mathrm{m}$ below an empirically estimated threshold value, and then confirmed by abrupt cooling, see corresponding images in Fig.~\ref{Fig2}~\textbf{c}. 
The expected position of the soliton is given by the coordinates of the  $\Omega_\mathrm{m}$, as depicted in Fig.\ref{Fig2}~\textbf{b}.
The CD nucleation is well reproduced in MC simulations~\cite{suppl}.

\begin{figure*}[ht]
\centering
\includegraphics[width=17cm]{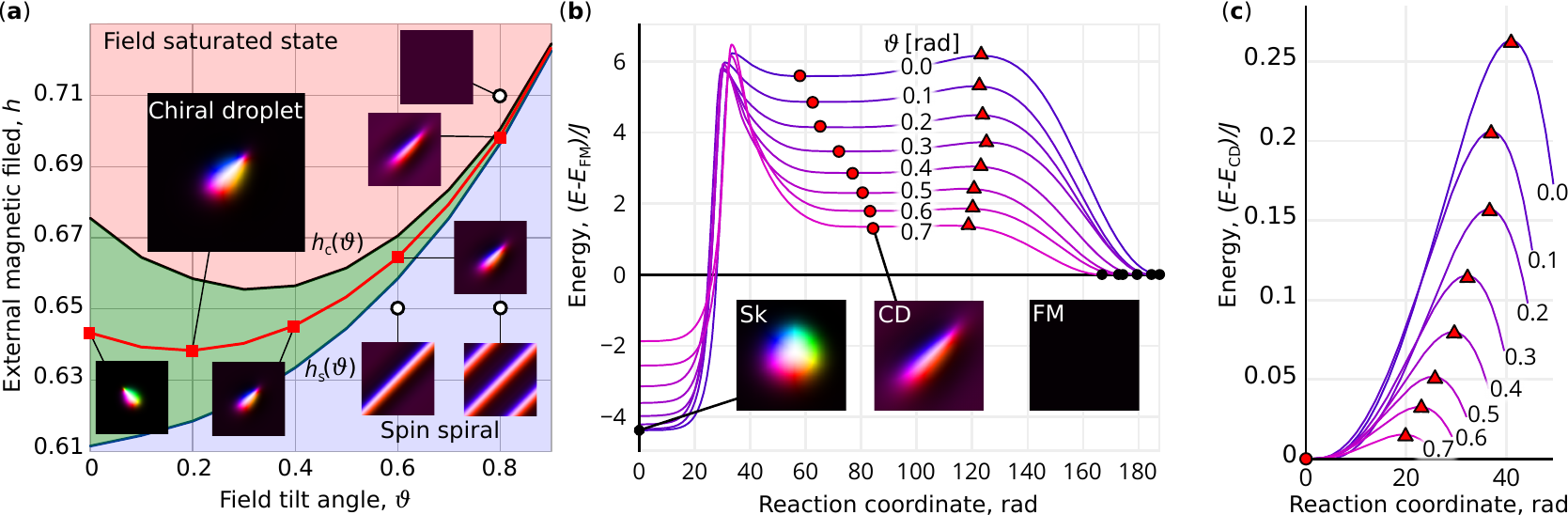}
\caption{\small
Stability diagram of CD in terms of the magnetic field, $h$, and its tilt, $\vartheta$. 
For $h<h_\mathrm{s}$ (blue region), the CD becomes unstable with respect to stretching.
%. 
For $h>h_\mathrm{c}$ (red region), the CD collapses to FM. 
The red line corresponds to the intermediate values of the stability range in terms of $h$; magnetic textures of CDs for some points on this curve are shown in insets.
\textbf{(b)} shows the energy variation along the MEPs connecting the Sk, CD and FM states for different parameters $h,\vartheta$ along the red line in (\textbf{a}). Red circles denote the energy minima corresponding to CD, red triangles mark the saddle points between CD and FM states. (\textbf{c}) shows the zoom of (\textbf{b}), with the zero of energy defined at the CD state.
}
\label{Fig3}
\end{figure*}

Figure~\ref{Fig3} \textbf{a} illustrates the stability range of CD in terms of absolute value of the applied magnetic field field, $h=|\mathbf{h}|$, and the tilt angle, $\vartheta$, for $T=0$. 
The CD stability domain is confined between the critical lines defined by the collapse field $h_\mathrm{c}(\vartheta)$, and by the stretching (or elliptical) instability field $h_\mathrm{s}(\vartheta)$: $h_\mathrm{s}(\vartheta)<h<h_\mathrm{c}(\vartheta)$. 
The range of $h$ 
where CD is stable shrinks with increasing $\vartheta$, and at $\vartheta\simeq 1.1$ the CD becomes unstable. 
Remarkably, the critical field $h_\mathrm{s}$ coincides with the phase transition line between skyrmion lattice and spin spiral states. 
Thereby, the CD always represents a metastable solution while the lowest energy state, in that range of fields, is the skyrmion lattice. 
Nevertheless, at moderate temperatures, the nucleation of the topologically trivial CDs dominates over the nucleation of topologically protected and energetically more favorable $\pi$-Sks. 
At elevated temperatures, we observe the transition into skyrmion lattice within an accessible simulation time irrespective of the field tilt angle, see Fig.~\ref{Fig1}. 

Further understanding of thermal nucleation of the CD states can be obtained using the harmonic transition state theory (HTST)~\cite{bessarab_2012,bessarab_2013}. 
Within the HTST, the rate of transition between states $X$ and $Y$ is described by the Arrhenius law,
\begin{align}
 & k^{X\rightarrow Y}=\nu^{X\rightarrow Y}\exp{\left(-\dfrac{\Delta E^{X\rightarrow Y}}{k_\mathrm{B}T}\right)},\label{Arr}
\end{align}
where the energy barrier $\Delta E^{X\rightarrow Y}$ can be identified from the MEP connecting $X$ and $Y$ as the energy difference between the highest point along the MEP -- the first-order saddle point (SP) on the energy surface of the system -- and the minimum at $X$. 
The pre-exponential factor $\nu^{X\rightarrow Y}$ incorporating the dynamical and entropic contributions to the transition rate is defined by the curvature of the energy surface at the minimum and at the SP~\cite{varentcova_2020}.

The MEP calculations using the geodesic nudged elastic band method~\cite{bessarab_2015,bessarab_2017} show that direct nucleation of the Sk state from the FM background is only possible for $\vartheta\lesssim 0.1$. For larger tilts of the external field, the MEP for the Sk nucleation and annihilation passes through an intermediate energy minimum corresponding to the CD state [see Fig.~\ref{Fig3}(b)]. Therefore, the system initially prepared in the FM state undergoes a transition to the CD state before it may reach the Sk state, which is also observed in the spin dynamics and Monte Carlo simulations~\cite{suppl}. As seen from Fig.~\ref{Fig3}\textbf{b}, the energy barrier between the FM state and CD state gradually decreases with $\vartheta$. 
This explains enhancement in the CD generation as the tilt of the field increases. 
On the other hand the energy of the saddle point between the CD and $\pi$-Sk states depends weakly on the tilt angle. 
It is mostly defined by the discretization of the system.
Approaching the micromagnetic regime with increasing  $L_\mathrm{D}$, the energy barrier $\Delta E^{\text{FM}\rightarrow \text{Sk}}$  increases~\cite{heil_2019}, which makes the $\pi$-Sk nucleation less probable.

\begin{figure}%[ht]
\centering
\includegraphics[width=8cm]{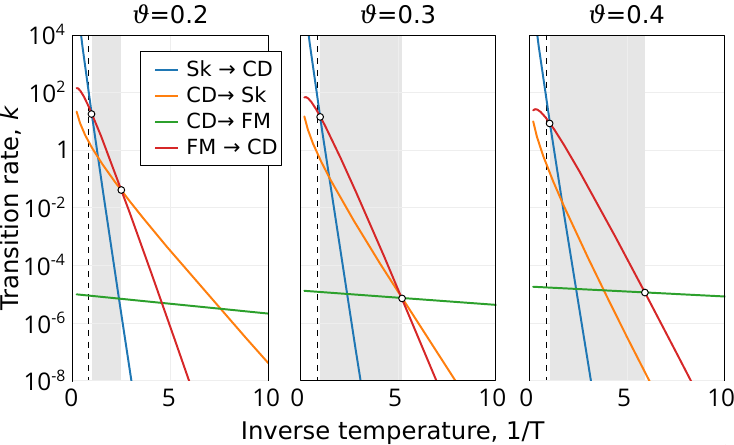}
\caption{\small 
Rates of various magnetic transitions (as indicated in the legend) as functions of the inverse thermal energy for various tilts, $\vartheta$, of the magnetic field. 
The amplitude of $h$ corresponds to the middle line of the CD stability range in Fig.~\ref{Fig3}\textbf{a}. The grey regions mark the temperature range with the maximal intensity of the FM$\rightarrow$CD transition. Vertical dashed line corresponds to the Curie temperature, $1/T_\mathrm{c}$.}
\label{mep}
\end{figure}

The CD energy minimum appears to be quite shallow [Fig.~\ref{Fig3}\textbf{b}], suggesting a quick collapse of the CD state. 
However, the HTST calculations, in agreement with the spin dynamics simulations, predict the opposite.
The rates of transitions involving the CD state (FM$\rightleftarrows$CD, CD$\rightleftarrows$Sk) are characterized by very different values of the energy barrier and pre-exponential factor, as can be seen from the Arrhenius plots shown in Fig.~\ref{mep}. 
In particular, the pre-exponential factor $\nu^{\text{FM}\rightarrow \text{CD}}$ for the FM$\rightarrow$CD transition -- the  nucleation of CDs -- is much larger than that for the backward transition.
Despite the very low energy barrier for the CD$\rightarrow$FM transition above a crossover temperature, the nucleation of CDs becomes more intensive than their annihilation into the FM state. 
For small tilts of the external field, the CD quickly transits into the Sk state. However, with increasing $\vartheta$ due to increasing energy barrier $\Delta E^{\text{CD}\rightarrow \text{Sk}}$ the CD$\rightarrow$Sk transition is progressively suppressed, see Fig.~\ref{Fig3}\textbf{b}.
The temperature range where the CD nucleation dominates the annihilation by several orders of magnitude increases with $\vartheta$, see gray color domain in Fig.~\ref{mep}.
The HTST calculations therefore provide a consistent interpretation of thermally induced creation of CDs observed in our spin dynamics and MC simulations.

In conclusion, we proposed a robust protocol for the creation of long-lived CDs by means of thermal fluctuations in a 2D chiral magnet under the tilted magnetic field.
The protocol takes advantage of the entropic stabilization and relatively low energy barrier for the nucleation of a topologically-trivial magnetic soliton. By varying the temperature and the tilt of the applied field, CDs and Sks can be generated selectively in a single system.
Co-existing CDs and Sks can further be used as building blocks for creating more complex magnetic solitons in chiral systems.

\begin{acknowledgments}
The authors would like to thank T. Sigurj\'onsd\'ottir for helpful discussions. This work was funded by Deutsche Forschungsgemeinschaft (DFG) through SPP 2137 "Skyrmionics" Grant No. KI 2078/1-1, the Russian Science Foundation (Grant No. 19-72-10138), the Icelandic Research Fund (Grant Nos. 184949 and 217750), and the University of Iceland Research Fund. 
\end{acknowledgments}
% \appendix

\end{document}